\documentstyle[aps]{revtex}
\newcommand{\R}{{\sf R\hspace*{-0.9ex}\rule{0.15ex}%
{1.5ex}\hspace*{0.9ex}}}
\begin{document}
\draft
%
%
\def \Ha #1 { {\cal H}_{ #1 } }
\def \Ho { {\Ha 1 } }
\def \Ht { {\Ha 2 } }
\def \M {M^{3n}}
\def \th {{(3)}}
\def \tw {{(2)}}
\def \a {{\alpha}}
\def \b {{\beta}}
\def \omegath {{\omega^{\th}}}
\def \omegatw {{\omega^{\tw}}}
\def \div #1 #2 {{ { #1 } \over { #2 } }}
\def \dbd #1 #2 { { \div {\partial { #1 }} {\partial { #2 }} }}
\def \pf {{\bf Proof: }}
\def \ex {{\bf Example: }}
\def \T {{\cal T}}
\def \X {{\cal X}}
\def \S {{\cal S}}
\def \g {{\bf g}}
\newtheorem{defn}{Definition}[section]
\newtheorem{thm}{Theorem}[section]
\newtheorem{prop}{Proposition}[section]
\newtheorem{lemma}{Lemma}[section]
\newtheorem{cor}{Corollary}[thm]
%
%
\title{ On generalized Nambu mechanics}
 \author{ Sagar A. Pandit\footnote{e-mail:
sagar@physics.unipune.ernet.in} and Anil D. Gangal\footnote{e-mail:
adg@physics.unipune.ernet.in}
 \\ \it Department of Physics, University of Poona, Pune 411 007,
 India.} \maketitle \begin{abstract} A geometric formulation of a
generalization of Nambu mechanics is proposed. This formulation is
carried out, wherever possible, in analogy with that of Hamiltonian
systems. In this formulation, a strictly nondegenerate constant 3-form
is attached to a 3n- dimensional phase space. Time evolution is
governed by two Nambu functions. A Poisson bracket of 2-forms is
introduced, which provides a Lie-algebra structure on the space of
2-forms. This formalism is shown to provide a suitable framework for
the descrip tion of non-integrable fluid flow such as the Arter flow,
the Chandrashekhar flow and of the coupled rigid bodies.
\end{abstract} 
%
%
\section { Introduction }

In 1973, Nambu proposed a generalization of Hamiltonian mechanics by
considering systems which obey Liouville theorem in 3 dimensional phase
space ~\cite{Nambu}.  In this formalism, the points of the phase space
were labeled by a canonical triplet $\tilde{r} = (x,y,z) $. A pair of
Hamiltonian like functions $H_1$, $H_2$, (which we call Nambu functions
hereafter), on this phase space were introduced. In terms of these
functions the equations of motion were written as \begin{equation} { {
d \tilde{r} } \over {dt} } = \tilde{\nabla} H_1 \times \tilde{\nabla}
H_2 \label{Nambuequation} \end{equation}

Nambu also defined a generalization of Poisson bracket on this new
phase space by
\begin{equation} \{ F, H_1, H_2 \} = \tilde{\nabla} F
\cdot ( \tilde{\nabla} H_1 \times \tilde{\nabla} H_2 )
\label{Nambubracket:0} 
\end{equation} 
An attempt was made by him to find a quantized version of the
formalism, but he succeeded just partially, since the correspondence
between the classical and quantum version is largely
lost~\cite{Nambu}.

The possibility of embedding the dynamics of a Nambu triplet in a four
dimensional canonical phase space formalism was proved
in~\cite{Mukunda,BandF} but such an embedding is local and non-unique.

An algebraic approach, which was suitable for quantization was
developed in ~\cite{KandT,Takht} where a generalization of the Nambu
bracket was postulated. In this approach a rather rigid consistency
condition, called the Fundamental identity, which is a generalization
of the Jacobi identity for Poisson brackets, was introduced. The
algebraic approach, no doubt, is quite elegant but is too restri ctive;
in the sense that the dynamics on a $n$-dimensional manifold is
determined by $(n-1)$ functions $H_1, \ldots,H_{n-1}$ which are
integrals of motions. Due to these large number of integrals of motion,
this formalism is not suitable for the formulati on of non-integrable
or chaotic systems.

Geometric formulation of Hamiltonian mechanics has revealed several
deep insights.  One would expect that a similar insight would emerge
from geometric formulation of Nambu systems. Such possibilities were
first examined by Estabrook~\cite{Esta} and more recently by
Fecko~\cite{Fecko}.

Recall that in the formulation of Hamiltonian systems, there exist
several equivalent ways~\cite{ArnoldIII} of endowing an even
dimensional manifold $M^{2n}$, with a symplectic structure. Two of the
prominent ways are:  \begin{enumerate} \item attach a closed
non-degenerate 2-form $\omega^{(2)}$.  \item attach a bracket on the
class of $C^\infty$ functions on $M^{2n}$ with properties of
bilinearity, skew-symmetry, Leibnitz rule, Jacobi identity and
non-degeneracy (i.e., Poisson structure together with non-degeneracy).
\end{enumerate}

Both types of approaches have been tried with Nambu
systems~\cite{Nambu,Mukunda,BandF,KandT,Takht,Esta,Fecko}
\begin{itemize} \item As mentioned above, it was found that the
Algebraic approach, starting with bracket of functions, required the
introduction of a rather rigid condition in the form of Fundamental
Identity~\cite{Takht,Okobu}, for consistency.  \item On the other hand
the geometric analysis~\cite{Esta,Fecko} led to the conclusion, that a
volume form was impossible to obtain from a 3-form, except in the most
trivial cases.  \end{itemize}

The complications mentioned above arise in both cases because the
volume preservation is a very stringent requirement. In light of these
findings we propose in this paper, that one need not impose volume
preservation for constructing geometric formalism of Nambu systems.

In this paper, we restrict ourselves to what we call a Nambu system of
order 3. Such a system has a 3n dimensional phase space and two Nambu
functions. We introduce a Nambu manifold as a 3n-dimensional manifold
$\M$ together with a constant 3-form $\omegath$ which is strictly
non-degenerate (The notion of non-degeneracy of 2-forms requires
modification. This modified notion is called strict non-degeneracy).
There is a natural generalization of Darboux basis corresponding to the
Hamiltonian systems. Equati ons of motion (Nambu equations) are
introduced in terms of two Nambu functions which are analogous to
Hamiltonian function in Hamiltonian dynamics. A novel feature of the
present paper is -- there is a natural way of introducing the
Nambu-Poisson bracket of 2-forms.  The form $\omegath$ is preserved in
the present approach which may be compared with the preservation of
$\omega^\tw$ for a Hamiltonian system. In Hamiltonian dynamics all
powers of $\omegatw$ are canonical invariants. However, in the present
formalism, since $\omegath$ is a form of an odd order, no conclusions
about canonical invariants can be obtained from preservation of
$\omegath$ itself.

The real justification for such a generalization can emerge from
applications to realistic physical systems and from better algorithmic
strategies. We demonstrate that the non-integrable Arter flow and the
Chandrashekhar flow describing Rayleigh -Benard convective motion with
rotation, can suitably be described in our framework. We further note
that the algorithmic strategy developed in~\ cite{HandK} can now be
identified as a generalization in the our framework of symplectic
integration corresponding to Hamiltonian framework.

In section~\ref{Geometricformulation} we have developed the geometric
formulation. In section~\ref{Nvectorspace}, The notion of strict
non-degeneracy of 3-forms is introduced. The notion of Nambu vector
space is defined using strict non-degeneracy. The
 existence of a Darboux like basis is proved. This is followed by the
 notion of a Nambu map. In section~\ref{Nmanifold} a Nambu manifold of
 order 3 and canonical transformations on this manifold are defined.
 Further a correspondence between 2-forms and vectors is established.
 This is followed by a discussion of conditions under which the 3-form
 $\omegath$ is preserved. In section~\ref{Nsystem} the Nambu systems
 are defined. A Nambu vector field corresponding to two Nambu functions
 is introduced. It is proved that the phase flow preserves the Nambu
 structure. In section~\ref{Nbracket} a bracket of 2-forms is
 introduced. This bracket provides a Lie algebra structure on the space
 of 2-forms.

In section~\ref{Applications} concrete applications of this framework
to the examples of coupled rigid bodies and to the fluid flows are
described.
%
%
\section { Geometric formulation of Nambu Systems }
\label{Geometricformulation}

We begin by recalling the essential features of the Hamiltonian
formalism.  The phase space has the structure of a smooth manifold $M$.
A closed, nondegenerate 2-form, viz the symplectic 2-form $\omegatw$,
is attached to this manifold. The non-degeneracy of $\omegatw$ imposes
the condition that $M$ be even dimensional. Canonical transformations
are those transformations under which the 2-form $\omegatw$ remains
invariant. Use of $\omegatw$ allows us to establish an isomorphism
between 1-forms and vector fields. The time evolution is governed by
the Hamiltonian vector field $X_H$ which is just the vector field
associated with the 1-form $dH$, where $H$ is a smooth function on
$M$.

Alternatively one can introduce Poisson brackets on the space of
$C^\infty$ functions on $M$. Poisson bracket together with
non-degeneracy condition induces a symplectic structure.

In the present paper, wherever possible, we develop the framework of
Nambu systems in analogy with that of Hamiltonian systems.
%
%
\subsection { Nambu vector space }
\label{Nvectorspace}

In this section we define the Nambu vector space which is analogous to
the symplectic vector space in Hamiltonian mechanics. Nambu vector
space is a vector space with a strictly non-degenerate 3-form. We prove
that in such a space there exists a preferred
 choice of basis which we call as Nambu-Darboux basis.

\begin{defn} (Nondegenerate form) : Let $E$ be a finite dimensional
vector space and let $\omegath$ be a 3-form on $E$ i.e.,
\begin{eqnarray} \omegath : E \times E \times E \rightarrow \R
\nonumber \end{eqnarray} the form $\omegath$ is called a
\underline{nondegenerate form} if \begin{eqnarray} \forall\;\;
non\;\;zero\;\;e_1 \in E\;\; \exists \;\; e_2, e_3 \in E\;\;
 such \;\; that \;\; \omegath (e_1, e_2, e_3) \not= 0 \nonumber
\end{eqnarray} \end{defn} {\bf Remarks:} \begin{enumerate} \item In
three dimensions every non zero 3-form is non-degenerate.  \item If the
dimension of the vector space is less than three, then one cannot have
an anti-symmetric, non-degenerate 3-form.  \item A nondegenerate 3-form
allows us to define an analog of orthogonal complement as follows.
\end{enumerate} \begin{defn} (Nambu complement) : Let $E$ be an $m$
dimensional vector space with $m \geq 3$. Let $\omegath$ be an
anti-symmetric and non-degenerate 3-form on $E$. Let us choose
$e_1,e_2,e_3 \in E$ such that $\omegath(e_1,e_2,e_3) \not= 0$. Let $P_1
= Spa n(e_1,e_2,e_3)$, then the \underline{Nambu complement} of $P_1$
is defined as \begin{eqnarray} P_1^{\perp_E} = \{ z \in E \;\;|\;\;
\omegath(z,z_1,z_2) = 0 \;\;\forall\;\;z_1,z_2 \in P_1 \} \nonumber
\end{eqnarray} \end{defn}

\begin{prop} Let $E$ be an $m$ dimensional vector space with $m \geq
3$. Let $\omegath$ be an anti-symmetric and non-degenerate 3-form on
$E$. Let us choose $e_1,e_2,e_3 \in E$ such that $\omegath( e_1,e_2,e_3
) \not= 0$. We further choose $e_2,e_3$ such that
$\omegath(e_1,e_2,e_3) = 1$. Let $P_1 = Span(e_1,e_2,e_3)$, then $E =
P_1 \oplus P_1^{\perp_E}$ \end{prop} \pf We write $\forall$ $x\in E$
\begin{eqnarray} x^\prime = x - \omegath(x,e_2,e_3) e_1
	     - \omegath(x,e_3,e_1) e_2 - \omegath(x,e_1,e_2) e_3
\nonumber \end{eqnarray} It is easy to see that $x^\prime \in
P_1^{\perp_E}$. From definition of $P_1^{\perp_E}$ it follows that $P_1
\cap P_1^{\perp_E} = \{ 0 \}$. Hence $E = P_1 \oplus P_1^{\perp_E}$
\newline\rightline{$\Box$}

\begin{defn} (Strictly nondegenerate form) : Let $E$ be an m
dimensional vector space and $\omegath$ be an anti-symmetric and
non-degenerate 3-form on $E$, the $\omegath$ is called
\underline{strictly non-degenerate} if for each non zero $e_1 \in E$
$\exists$ a two dimensional subspace $E_1 \subset E$ such that
\begin{enumerate} \item $\omegath(e_1,x_1,x_2) \not= 0$ $\forall$
linearly independent $\{e_1,x_1,x_2\}$ where $x_1,x_2 \in F_1$ and $F_1
= Span(e_1 + E_1)$.  \item $\omegath(e_1,z_1,z_2)=0$ $\forall$ $z_1,
z_2 \in F_1^{\perp_E}$.  \end{enumerate} \end{defn} {\bf Remarks:}
\begin{enumerate} \item In three dimensional space every non-degenerate
form is strictly non-degenerate.  \end{enumerate}

\begin{defn} (Nambu vector space) : Let $E$ be a finite dimensional
vector space and $\omegath$ be a completely anti-symmetric and strictly
nondegenerate 3-form on $E$ then the pair $(E,\omegath)$ is called a
\underline{Nambu vector space}.  \end{defn}

Recall that the rank of 2-form is defined as the rank of its matrix
representation. We now introduce the notion of rank for the
anti-symmetric 3-forms.  \begin{defn} (Rank of $\omegath$) : Let $E$ be
a finite dimensional vector space and $\omegath$ be a completely
anti-symmetric 3-form, then the \underline{rank of $\omegath$} is
defined as \begin{eqnarray} \sup_{P \subset E} \{ d\; |\; d = dim P,\;
(P,\omegath|_P)\; is\;\;a\;\; Nambu \;\;vector \;\;space\; \} \nonumber
\end{eqnarray} \end{defn} {\bf Remarks:} \begin{enumerate} \item The
following proposition gives the prescription to construct the
Nambu-Darboux basis.  \end{enumerate}

\begin{prop} \label{NDthmvecspace} Let $E$ be an m dimensional vector
space. Let $\omegath$ be a 3-form of rank $m$ on $E$, then $m=3n$ for a
unique integer $n$. Further there is an ordered basis $\{ e_i \}, i =
1,\ldots,m$ with the corresponding dual basis $\{ \alpha^i \},
i=1,\ldots,m$, such that \begin{eqnarray} \omegath &=& \sum_{i=0}^{n-1}
\alpha^{3i+1} \wedge \alpha^{3i+2} \wedge \alpha^{3i+3}\;\;\; if \;\;n
> 0  \nonumber \\ \omegath &=& 0 \;\;\; otherwise \nonumber
\end{eqnarray} \end{prop} \pf The rank of $\omegath$ is $m$, implies
that $(E,\omegath)$ is a Nambu vector space. One can assume that $m
\geq 3$, for otherwise the proposition is trivial with $n=0$. Let $e_1
\in E$ and let $E_1 \subset E$ be a 2-dimensional subspace such that
$\omegath (e_1, e_2, e_3) \not= 0$ $\forall$ linearly independent
$\{e_1,e_2,e_3\}$ where $e_2, e_3 \in P_1$ and $P_1 = Span(e_1 + E_1)$.
Let $e_2, e_3$ be a basis of $E_1$, then $e_1,e_2,e_3$ is a basis of
$P_1$.  Thus $(P_1,\omegath |_{P_1})$ is three dimensional Nambu vector
space.  It follows that one can write $\omegath|_{P_1}$ in basis $\a^1,
\a^2, \a^3$ dual to $e_1,e_2,e_3$ as \begin{eqnarray} \omegath |_{P_1}
= \alpha^1 \wedge \alpha^2 \wedge \alpha^3 \nonumber \end{eqnarray} If
$m=3$ then $P_1 = E$ and the proposition is proved with $n=1$. Hence we
assume $m > 3$. We denote $Q_1 = P_1^{\perp_E}$.  The dimension of
$Q_1$ is $m-3$. Since the vector space $E$ is Nambu vector space it
follows $dim (Q_1) \geq 3$.  Let $e_4 \in Q_1 \subset E$ $\Rightarrow$
$\exists$ $e_5, e_6 \in E$ such that $\omegath(e_4, e_5, e_6) \not= 0$
$\forall$ linearly independent $\{e_4, e_5, e_6\}$ where $e_5, e_6 \in
P_2$ and $P_2 = Span (e_4, e_5, e_6)$. By definition of $Q_1$ it follow
s that $e_5,e_6 \not\in P_1$. We choose $e_5, e_6 \in P_1^{\perp_E} =
Q_1$. It follows from the strict non-degeneracy of $\omegath$ in $E$
and the facts $P_2^{\perp_{Q_1}} \subset P_2^{\perp_E}$ and $P_2
\subset Q_1$ that $(Q_1, \omegath |_{Q_1})$ is a Nambu vector space of
dimension $m-3$.

repeated application of the above argument on $Q_1$ in place of $E$ and
so on yields \begin{eqnarray} E = P_1 \oplus \cdots \oplus P_n
\nonumber \end{eqnarray} We stop the recursion when $dim(Q_n) = 0$.
Using the strict non-degeneracy of $\omegath$ we get \begin{eqnarray}
\omegath &=& \omegath |_{P_1} + \cdots + \omegath |_{P_n} \nonumber \\
&=& \sum_{i=0}^{n-1} \a^{3i+1} \wedge \a^{3i+2} \wedge \a^{3i+3}
\nonumber \end{eqnarray} \newline \rightline {$\Box$}

\begin{defn} (Nambu mappings) : If $(E,\omega)$ and $(F,\rho)$ are two
Nambu vector spaces and $f : E \rightarrow F$ is a linear map such that
the pullback $f^* \rho = \omega$, then $f$ is called a \underline{Nambu
mapping}.  \end{defn} \begin{prop} Let $(E,\omega)$ and $(F,\rho)$ be
two Nambu vector spaces of dimension $3n$ and let a linear map $f: E
\rightarrow F$ be Nambu mapping then $f$ is an isomorphism on the
vector spaces.  \end{prop} \pf Let if possible, $f$ be singular. Then
there exists $x \in E$ and $x \not= 0$ such that $fx = 0$. But since
$f$ is a Nambu mapping one can write \begin{eqnarray}
\rho(f(x),f(y),f(z)) = \omega(x,y,z) \nonumber \end{eqnarray} where
$y,z \in E$ are so chosen that $\omega(x,y,z) \not= 0$. Which leads to
contradiction. Hence $f$ is an isomorphism.  \newline \rightline
{$\Box$} \begin{prop} Let $(E,\omegath)$ be a Nambu vector space of
dimension $3n$, then the set of Nambu mappings from $E$ to $E$ forms a
group under composition.  \end{prop} \pf Now let $f$ and $g$ are Nambu
mappings.  \begin{eqnarray} (f \circ g )^* \omega^\th = g^* \circ f^*
\omega^\th = g^* \omega^\th = \omega^\th \nonumber \end{eqnarray} and
\begin{eqnarray} ( f^{-1})^* \omega^\th = (f^*)^{-1} \omega^\th =
\omega^\th \nonumber \end{eqnarray} \newline \rightline {$\Box$}
%
%
\subsection { Nambu manifold }
\label{Nmanifold}

In analogy with the notion of symplectic manifold we now introduce
Nambu manifold.  \begin{defn} (Nambu Manifold) : Let $\M$ be a
$3n$-dimensional $C^\infty$ manifold and let $\omegath$ be a 3-form
field on $\M$ such that $\omegath$ is completely anti-symmetric,
constant (i.e., a constant section on the bundle of 3-forms) and
strictly nondegenerate at every point of $\M$ then the pair
$(\M,\omegath)$ is called
 a \underline{Nambu manifold}.  \end{defn} {\bf Remarks:}
\begin{enumerate} \item In the case of Hamiltonian systems the form
$\omegatw$ is assumed to be closed, here $\omegath$ is assumed to be a
constant form. The condition of closedness allows many 3-forms which in
general may not be consistent with the non-degeneracy conditio n.
\end{enumerate} \begin{thm} (Nambu-Darboux theorem) : \label{NDthm} Let
$(\M,\omegath)$ be a Nambu manifold then  at every point $p \in \M$,
there is a chart $(U,\phi)$ in which $\omegath$ is written as
\begin{eqnarray} \omegath |_U = \sum_{i=0}^{n-1} dx_{3i+1} \wedge
dx_{3i+2} \wedge dx_{3i+3} \nonumber \end{eqnarray} where
$(x_1,x_2,x_3,\ldots,x_{3(n-1)+1},x_{3(n-1)+2},x_{3(n-1)+3})$ are local
coordinates on $U$ described by $\phi$.  \end{thm} \pf Proof follows
from Proposition ~\ref{NDthmvecspace}.  \newline \rightline {$\Box$}

The coordinates described in theorem ~\ref{NDthm} will be called
Nambu-Darboux coordinates hereafter. We use these coordinates in the
remaining parts of the paper.

\begin{defn} (Canonical transformation) : Let $(\M,\omegath)$ and
$(N^{3n},\rho^\th)$ be Nambu manifolds. A $C^\infty$ mapping $F : \M
\rightarrow N^{3n}$ is called \underline{canonical transformation} if
$F^*\rho^\th = \omegath$.  \end{defn}

Let $\T^0_k (\M)$ denote a bundle of k-forms on $\M$, $\Omega^0_k (\M)$
denote the space of k-form fields on $\M$ and $\X(\M)$ denote the space
of vector fields on $\M$. Now for a given vector field $X$ on $\M$ we
denote \begin{eqnarray} i_X : \Omega^0_k (\M) \rightarrow
\Omega^0_{k-1} (\M) \nonumber \end{eqnarray} as inner product of $X$
with k-form or contraction of a k-form by $X$ given by \begin{eqnarray}
(i_X \eta^{(k)}) (a_1,\ldots,a_{k-1}) = \eta^{(k)} (X,a_1, \ldots,
a_{k-1}) \nonumber \end{eqnarray} where $\eta^{(k)} \in \Omega^0_k(\M)$
and $a_1, \ldots, a_{k-1} \in \X(\M)$

We define now the analogs of raising and lowering operations.  The map
$\flat : \X(\M) \rightarrow \Omega^0_2(\M)$ is defined by $ X \mapsto
X^\flat = i_X \omegath$.  Whereas the map $\sharp : \Omega^0_2(\M)
\rightarrow \X(\M)$, is defined by the following prescription.  Let
$\alpha$ be a 2-form and $\alpha_{ij}$ be its components in
Nambu-Darboux coordinates, then the components of $\alpha^\sharp$ are
given by \begin{eqnarray} \label{twoformtovec} {\alpha^\sharp}^{3i+p} =
{1 \over 2} \sum_{l,m=1}^3 \varepsilon_{plm} \alpha_{3i+l \;\;\; 3i+m}
\end{eqnarray} where $0 \leq i \leq n-1$, $ p=1,2,3$ and
$\varepsilon_{plm}$ is the Levi-Cevit\`a symbol. \\ {\bf Remarks:}
\begin{enumerate} \item It may appear that components of $\a^\sharp$
have been given a definition using a particular choice of coordinate
system. The definition itself is actually coordinate free as shown in
the Appendix.

\item In difference with the customary meaning of $\flat$ and $\sharp$
used in ordinary tensor analysis we note that in this paper $\flat$
maps a vector to 2-form and not to a 1-form, also $\sharp$ maps a
2-form to a vector. From the above definition it is clear that
$(X^\flat)^{\sharp} = X$ but $(\alpha^{\sharp})^\flat$ may not always
yield the same $\alpha$.

\item In fact consider $\T^0_{2_x}(\M)$, the space of 2-forms at $x \in
\M$. The $\sharp$ defines an equivalence relation on $\T^0_{2_x}(\M)$
as follows. Let $\omega_1^\tw(x), \omega_2^\tw (x) \in \T^0_{2_x}(\M)$.
We say that $ \omega_1^\tw(x) \sim \omega _2^\tw (x)$ if
$(\omega_1^\tw)^\sharp (x) = (\omega_2^\tw)^\sharp (x)$.  It is easy to
see that $\sim$ is an equivalence relation.  We define the equivalence
classes $\S^0_{2_x}(\M) = \T^0_{2_x}(\M) / \sim$.

Let $\omega_1^\tw(x), \omega_2^\tw(x), \omega_3^\tw(x) \in
\T^0_{2_x}(\M)$ and the equivalence class be denoted by $[\;]$ i.e.,
$\a_1(x) = [\omega_1^\tw(x)],\a_2(x) = [\omega_2^\tw(x)], \a_3(x) =
[\omega_3^\tw(x)]$ where $\a_1(x), \a_2(x), \a_3(x) \in \S^
0_{2_x}(\M)$. The addition and scalar multiplication on
$\S^0_{2_x}(\M)$ are defined as follows.  \begin{eqnarray} \a_1(x) +
\a_2(x)&=& [ \omega_1^\tw (x) + \omega_2^\tw (x) ] \nonumber \\ \mu
\cdot \a_1(x) &=& [ \mu \cdot \omega_1^\tw (x) ] \nonumber
\end{eqnarray} where $\mu \in \R$. It is easy to see that
$(\S^0_{2_x}(\M),+,\R,\cdot)$ forms a vector space and the dimension of
this vector space is $3n$.  \end{enumerate}

Now we investigate conditions under which the given Nambu form
$\omegath$ is invariant under the action of the vector field
$\beta^\sharp$ associated with any 2-form $\beta$.

\begin{prop}\label{equivofvec} Let $\beta^{(2)} \in \Omega^0_2(\M)$
then $(\beta^{(2)^\sharp})^\flat \sim \beta^{(2)}$ \end{prop} \pf Proof
follows from the fact that $(X^\flat)^\sharp = X \;\; \forall X \in
\X(\M)$ \newline \rightline {$\Box$}

\begin{thm}\label{equivcloseform} Let $\beta^{(2)} \in \Omega^0_2(\M)$,
and $f^t$ be a flow corresponding to $\beta^{(2)^\sharp}$, i.e., $f^t :
\M \rightarrow \M$ such that \begin{eqnarray} {\div d { dt } }
\Big|_{t=0} ( f^tx ) = (\beta^{(2)^\sharp}) x \;\; \forall x \in \M
\nonumber \end{eqnarray} Then the form $\omegath$ is preserved under
the action of $\beta^{(2)^\sharp}$ iff $d(\beta^{(2)^\sharp})^\flat =
0$.  i.e., ${f^t}^* \omegath = \omegath$ iff
$d(\beta^{(2)^\sharp})^\flat = 0$ \end{thm} \pf \begin{eqnarray} {\div
d {dt} } ( {f^t}^* \omegath ) &=& {f^t}^* ( L_{\beta^{(2)^\sharp}}
\omegath ) \nonumber \\ &=& {f^t}^* ( i_{\beta^{(2)^\sharp}} d\omegath
+ d (i_{\beta^{(2)^\sharp}} \omegath)) \nonumber \\ &=& {f^t}^*
d(\beta^{(2)^\sharp})^\flat \nonumber \end{eqnarray} If
$d(\beta^{(2)^\sharp})^\flat = 0 \Rightarrow {\div d {dt} } ( {f^t}^*
\omegath ) = 0$ and also if ${\div d {dt} } ( {f^t}^* \omegath ) = 0
\Rightarrow d(\beta^{(2)^\sharp})^\flat = 0$ \newline\rightline{$\Box$}
From the above theorem it follows that the vector field corresponding
to a 2-form preserves Nambu structure if that 2-form is equivalent to a
closed 2-form. By Poincar\'e lemma one can locally write the closed
2-form as $d\xi$ where $\xi$ is 1-form. Witho ut loss of generality we
choose $\xi = f_1 d f_2$ where $f_1$ and $f_2$ are $C^\infty$ functions
on $\M$. So we can choose $(\beta^{\tw^\sharp})^\flat$ as $df_1 \wedge
df_2$ and by proposition ~\ref{equivofvec} $(df_1 \wedge df_2) \sim
\beta^\tw$.
%
%
\subsection { Nambu system }
\label{Nsystem}

Having introduced the relevant structure viz. the Nambu manifold, we
now proceed with the discussion of time-evolution. The
time-time-evolution is governed by a vector field obtained from two
Nambu functions.  \begin{defn} (Nambu vector field) : Let ${\Ha 1 }$,
${\Ha 2 }$ be real valued $C^\infty$ functions (Nambu functions) on
$(\M,\omegath)$ then $N$ is called \underline{Nambu vector field}
corresponding to ${\Ha 1 }, {\Ha 2 }$ if \begin{eqnarray} N = (d {\Ha 1
} \wedge d {\Ha 2 })^\sharp \nonumber \end{eqnarray} \end{defn}

\begin{defn} (Nambu system) : A four tuple $(\M,\omegath, {\Ha 1 },
{\Ha 2 })$ is called \underline{Nambu system}.  \end{defn} Henceforth
we choose $d\Ho \wedge d\Ht$ as the representative 2-form of the class
$[d\Ho \wedge d\Ht]$. \footnote{ We note that in the equivalence class
of $d\Ho \wedge d\Ht$ there are some 2-forms which are not closed and
can not be expressed as $d\Ho \wedge d\Ht$. }

Now for a given representative 2-form we have the freedom in the choice
of $\Ho$ and $\Ht$. This freedom is discussed in ~\cite{Nambu} as gauge
freedom in the choice of $\Ho$ and $\Ht$. \footnote{ Note that this
freedom is different from the freedom in de finition of $\S^0_2$. This
freedom is generalization of the freedom of additive constant in
Hamiltonian dynamics.} \begin{defn} (Nambu phase flow) : Let $(\M,
\omegath , {\Ha 1 }, {\Ha 2 })$ be a Nambu system then the
diffeomorphisms $g^t : \M \rightarrow \M$ satisfying \begin{eqnarray}
{\div d {dt} } \Big|_{t=0} (g^t {\bf x}) &=& (d{\Ha 1 } \wedge d {\Ha 2
})^\sharp {\bf x} \;\;\; \forall\; {\bf x} \in \M \nonumber \\ &=& N
{\bf x} \nonumber \end{eqnarray} is called \underline{Nambu phase
flow}.  \end{defn}

From the properties of flow of a differentiable vector field it follows
that $g^t$ is one parameter group of diffeomorphisms.

\begin{thm} \label{Preserveomegaunderg} Nambu phase flow preserves
Nambu structure. i.e., \begin{eqnarray} g^{t*} \omegath = \omegath
\nonumber \end{eqnarray} \end{thm} \pf Proof follows from
Theorem~\ref{equivcloseform} \newline \rightline {$\Box$}

{\bf Remarks:} \begin{enumerate} \item Since the proof of
theorem~\ref{Preserveomegaunderg} is valid for any time $t$ provided
the flow $g^t$ exists, it is automatically valid for any time interval
say from $t_1$ to $t_2$. Further the flow preserves the $\omegath$,
which implies that the map $g^t : \M \rightarrow \M$ is a canonical
transformation. This leads to the interpretation, as in the case of
Hamiltonian systems viz,:{\it `` The history of a Nambu system is a
gradual unfolding of successive canonical transformation.''} Such an o
bservation is one of the crucial ingredients required for the
development of symplectic integrators for Hamiltonian systems. The
present observation therefore can be used for a similar algorithm for
Nambu systems.  \end{enumerate} \begin{prop} Let
$(\M,\omegath,\Ho,\Ht)$ be a Nambu system, then $\Ho$ and $\Ht$ are
constants of motion. \label{ConstH} \end{prop} \pf We prove the result
for $\Ho$ and proof is similar for $\Ht$ \begin{eqnarray} {\div d {dt}
} {\Ho} &=& L_N \Ho \nonumber\\ &=& i_N d\Ho \nonumber\\ &=& d\Ho(N)
\nonumber\\ &=& d\Ho((d\Ho \wedge d\Ht)^\sharp) \nonumber
\end{eqnarray} If we write the RHS in Nambu-Darboux coordinates using
equations (~\ref{twoformtovec}) we get RHS = 0.  \newline
\rightline{$\Box$}
%
%
\subsection { Nambu bracket }
\label{Nbracket}

We now define the analog of the Poisson bracket for 2-forms. This leads
to the algebra of 2-forms. Further we also define the brackets for
three functions in the conventional fashion~\cite{Nambu}.  \begin{defn}
(Nambu Poisson bracket) : Let $\omega_a^\tw$ and $\omega_b^\tw$ be
2-forms then the \underline{Nambu Poisson bracket} is a map $\{,\} :
\Omega^0_2(\M) \times \Omega^0_2(\M) \rightarrow \Omega^0_2(\M)$ given
by \begin{eqnarray} \{ \omega_a^\tw , \omega_b^\tw \} = [
{\omega_a^\tw}^\sharp , {\omega_b^\tw}^\sharp ]^\flat \nonumber
\end{eqnarray} where $[,]$ is Lie bracket of vector fields.  \end{defn}
that is, one writes \begin{eqnarray} \{\omega_a^\tw, \omega_b^\tw \}
(\xi,\eta) = (i_{[{\omega_a^\tw}^\sharp, {\omega_b^\tw}^\sharp ]}
\omegath) (\xi, \eta) \;\;\;\; \forall \;\; \xi, \eta \in \X(\M)
\nonumber \end{eqnarray}

From the definition it clear that if $\omega_a^\tw \sim \omega_b^\tw$
then the $\{\omega_a^\tw, \omega_b^\tw\} = 0 $ and if $\omega_a^\tw
\sim \omega_{a^\prime}^\tw$ and $\omega_b^\tw \sim
\omega_{b^\prime}^\tw$ then $\{ \omega_a^\tw, \omega_b^\tw \} = \{
\omega_{a^\prime}^\tw, \omega_{b^\prime}^\tw \} $

\begin{prop} \label{NBprop1} Let $(\M,\omegath)$ be a Nambu manifold
and $\a,\b \in \Omega^0_2(\M)$ then \begin{eqnarray} \{ \a, \b \} =
L_{\a^\sharp} {(\b^\sharp)^\flat} - L_{\b^\sharp} {(\a^\sharp)^\flat} -
d ( i_{\a^\sharp} i_{\b^\sharp} \omegath )  \nonumber \end{eqnarray}
\end{prop} \pf \begin{eqnarray} \{ \a, \b\} &=& i_{[\a^\sharp,
\b^\sharp]} \omegath \nonumber \\ &=& L_{\a^\sharp} ( i_{\b^\sharp}
\omegath ) - i_{\b^\sharp}(L_{\a^\sharp} \omegath) \nonumber \\
&=&L_{\a^\sharp} (\b^\sharp)^\flat - i_{\b^\sharp} d(i_{\a^\sharp}
\omegath) \nonumber \\ &=& L_{\a^\sharp} (\b^\sharp)^\flat -
L_{\b^\sharp} (\a^\sharp)^\flat - d(i_{\a^\sharp} i_{\b^\sharp}
\omegath ) \nonumber \end{eqnarray} \newline \rightline {$\Box$}

\begin{prop} Let $\a$ and $\b$ be 2-forms. Further let $\a^\sharp$ be a
Nambu vector field. Let $\a^\prime = (\a^\sharp)^\flat$ and $\b^\prime
= (\b^\sharp)^\flat$ then \begin{eqnarray} \{ \a, \b \} =
L_{{\a^\prime}^\sharp} {\b^\prime} \nonumber \end{eqnarray} \end{prop}
\pf Proof follows from proposition~\ref{NBprop1} \newline \rightline
{$\Box$}

\begin{prop} The space $\Omega^0_2(\M)$ forms a Lie Algebra with
multiplication defined by the bracket i.e., If $\alpha, \beta, \gamma
\in \Omega^0_2(\M)$ \begin{enumerate} \item $\{\alpha + \gamma, \beta\}
= \{ \alpha, \beta\} + \{\gamma, \beta \}$ and $\{\alpha, \beta +
\gamma \} = \{\alpha, \beta\} + \{\alpha, \gamma\}$ \item
$\{\alpha,\alpha\} = 0$ \item $\{\alpha,\{\beta, \gamma\}\} +
\{\beta,\{\gamma,\alpha\}\} + \{\gamma,\{\alpha, \beta\}\} = 0$
\end{enumerate} \end{prop} \pf Follows from definition of bracket
\newline \rightline{$\Box$} {\bf Remarks:} \begin{enumerate} \item In
the Hamiltonian systems smooth functions on the phase space are
considered as observables. To each such function $f$ a natural vector
field (viz $X_f$ which is in correspondence with $df$) is attached. We
note that it is really to the 1-form $df$
 that a vector field is attached (All functions differing by constants
 form an equivalence class producing identical $df$.)

In view of the above discussion it is clear that in the Nambu framework
2-forms play a basic role. In this connection we point out the
following facts:  \begin{enumerate} \item Vector fields are naturally
associated with 2-forms.  \item The bracket of 2-forms provides the Lie
Algebra structure. On the other hand bracket of functions introduces a
non-associative structure as noted by~\cite{Nambu,Okobu}.
\end{enumerate} \end{enumerate} \begin{defn} (Nambu bracket for
functions) : Consider a Nambu manifold $(\M, \omegath)$ and let $f,g,h$
be $C^{\infty}$ functions on $\M$ then \underline{Nambu bracket for
functions} is given by \begin{eqnarray} \{f,g,h \} = L_{(dg \wedge
dh)^\sharp}f = i_{(dg \wedge dh)^\sharp} df \nonumber \end{eqnarray}
\end{defn} By using equation (~\ref{twoformtovec}) in Nambu-Darboux
coordinates, \begin{eqnarray} \{ f, g, h \} = \sum_{i=0}^{n-1}
\sum_{k,l,m=1}^3 \varepsilon_{klm}
	     { \dbd {f} {x_{3i+k}} } { \dbd {g} {x_{3i+l}} } { \dbd {h}
	     {x_{3i+m}} }
\nonumber \end{eqnarray} In three dimensions this is simply the
definition of the bracket given by Nambu~\cite{Nambu} \begin{prop}
\label{NBproptwo} Consider a Nambu system $(\M, \omegath, g, h)$. Let
$f,g^\prime, h^\prime \in C^{\infty}(\M)$ satisfying $(dh \wedge dg)
\sim (dh^\prime \wedge dg^\prime)$ and $(df \wedge
dg^\prime)^{\sharp^\flat} = (df \wedge dg^\prime)$, then we have
\begin{eqnarray} \{f,g,h\} dg^\prime = i_{(dg \wedge dh)^\sharp} i_{(df
\wedge dg^\prime)^\sharp} \omegath \nonumber \end{eqnarray} \end{prop}
\pf \begin{eqnarray} \{ f, g, h \}dg^\prime &=& ( L_{(dg \wedge
dh)^\sharp} f ) dg^\prime \nonumber \\ &=& (i_{(dg\wedge dh)^\sharp}
(df \wedge dg^\prime)) \nonumber \\ &=& i_{(dg \wedge dh)^\sharp}
i_{(df \wedge dg^\prime)^\sharp} \omegath \nonumber \end{eqnarray}
\newline \rightline{$\Box$}

Following proposition gives the relation between the bracket of 2-forms
and the bracket of functions.  \begin{prop}\label{bracketrelation} Let
$(\M,\omegath)$ be a Nambu manifold and let $f,g,h_1,h_2$ be $C^\infty$
functions satisfying $(df \wedge dg)^{\sharp^\flat} = df \wedge dg$ and
$(dh_1 \wedge dh_2)^{\sharp^\flat} = dh_1 \wedge dh_2$ then
\begin{eqnarray} \{ dh_1 \wedge dh_2, df \wedge dg \} = d \{f,h_1,h_2\}
\wedge dg + df \wedge d\{g,h_1,h_2\} \nonumber \end{eqnarray}
\end{prop} \pf From Proposition ~\ref{NBprop1} we have \begin{eqnarray}
\{ dh_1 \wedge dh_2, df \wedge dg \} = L_{(dh_1 \wedge dh_2)^\sharp}
(df \wedge dg) \nonumber \end{eqnarray} \newline \rightline{$\Box$}

{\bf Remark :} \begin{enumerate} \item If a function $f$ is an integral
of motion then the Nambu bracket of function $\{f,\Ho,\Ht\}$ is zero
and conversely. On the other hand if $\b$ is a 2-form such that
$\{d\Ho\wedge\Ht,\b\} = 0$ then there exists a 2-form $\b^\prime$ in
the equivalenc e class of $\b$ which is an invariant of motion. Also by
proposition~\ref{bracketrelation} these two statements are consistent.
\end{enumerate}
%
%
\section { Applications of Nambu mechanics }
\label{Applications}

The purpose of this section is to demonstrate that there are systems
that can be described appropriately using the formalism developed
here.
%
%
\subsection { Fluid flows }

It was known for a long time that two dimensional incompressible fluid
flows can be studied using the two dimensional Hamiltonian framework.
It was realized by Holm and Kimura~\cite{HandK} that for three
dimensional integrable flows of incompressible fl uids in the
Lagrangian picture, Nambu description is suitable. However the three
dimensional Nambu system is not suitable as a framework for the
formulation of non-integrable fluid flows. We now claim that this
requirement can be fulfilled by an appropria te choice of $\Ho$, $\Ht$
in a $3n$ dimensional Nambu framework. Specifically we show below that
the Arter and Chandrashekhar flows (describing Rayleigh-Benard
convective motion) can be casted as flows on an invariant three
dimensional subspace of a six d imensional Nambu system.

We begin with the case of Chandrashekhar flow.
%
%
\subsubsection{ Chandrashekhar flow }

We now show that the Chandrashekhar flow admits the description in our
framework as follows: \\ Consider a Nambu manifold $(\R^6,\omegath)$.
Let $\{ x, y, z, x^\prime, y^\prime , z^\prime \} $  be the coordinates
on $\R^6$ and $\omegath = dx \wedge dy \wedge dz + dx^\prime \wedge
dy^\prime \wedge dz^\prime$ in these coordinates. Let ${\Ha 1 }$ and
${\Ha 2 }$ be the Nambu functions defined as \begin{equation} {\Ha 1 }
= \log \Big({ \sin(x) \over \sin(y)} \Big)
	 -  \log \Big({ \sin(x^\prime) \over \sin(y^\prime)} \Big) -
	K^2 { \cos(x^\prime) \over \sin(x^\prime) } ( y - y^\prime) -
	K^2 { \cos(y^\prime) \over \sin(y^\prime) } ( x - x^\prime)
\end{equation} \begin{equation} {\Ha 2 } = \sin(x) \sin(y) \sin(z) -
\sin(x^\prime) \sin(y^\prime)
	  \sin(z^\prime) + ( y - y^\prime )A + (x - x^\prime)B
\end{equation} where $K^2$ is constant which will be identified later,
and \begin{eqnarray} \nonumber A = { { K^2 \cos(y^\prime) \cos(y)
\sin^2(x) \sin(z) } \over { \cos(x) \sin(y^\prime) - K^2 \sin(x)
\cos(y^\prime) } } \end{eqnarray} \begin{eqnarray} B = - { { K^2
\cos(x^\prime) \cos(x) \sin^2(y) \sin(z) } \over { \cos(y)
\sin(x^\prime) + K^2 \sin(y) \cos(x^\prime) } } \nonumber
\end{eqnarray} We will now show that the sub-space defined by
\begin{equation} x = x^\prime, y = y^\prime, z = z^\prime
\label{sectplane} \end{equation} is an invariant subspace\footnote{
This subspace is not Nambu with respect to the $\omegath$ defined above
because the transformation from $\R^6$ to this subspace is not
canonical} and also that the equations of motion for $(x,y,z)$  ( or
equivalently $(x ^\prime, y^\prime, z^\prime)$ ) are precisely the
equations governing the Chandrashekhar flow ~\cite{HandK}.

We write the Nambu vector field for given functions ${\Ha 1 }$ and
${\Ha 2 }$ \begin{eqnarray} \dot x &=& - \sin(x) \cos(y) \cos(z) - K^2
\cos(x^\prime) \sin(y) \cos(z) { \sin(x) \over \sin(x^\prime) }
\nonumber \\
 &  & + \Big( - {\cos(y) \over \sin(y)} - K^2 {\cos(x^\prime) \over
 \sin(x^\prime) } \Big) \Big( { {\partial A} \over {\partial z} } ( y -
y^\prime) + { {\partial B} \over {\partial z} } ( x - x^\prime) \Big)
\\ \dot y &=& - \cos(x) \sin(y) \cos(z) + K^2 \cos(y^\prime) \sin(x)
\cos(z) { \sin(y) \over \sin(y^\prime) } \nonumber \\
 &  & - \Big( {\cos(x) \over \sin(x)} - K^2 {\cos(y^\prime) \over
 \sin(y^\prime) } \Big) \Big( { {\partial A} \over {\partial z} } ( y -
y^\prime) + { {\partial B} \over {\partial z} } ( x - x^\prime) \Big)
\\ \dot z &=& 2 \cos(x) \cos(y) \sin(z) + \Big( {\cos(x) \over
\sin(x)}  - K^2 { \cos(y^\prime) \over \sin(y^\prime) } \Big) \Big( {
{\partial A} \over {\partial y} } ( y - y^\prime )  + { {\partial B}
\over {\partial y} } (x -x^\prime) \Big) \nonumber \\
 &  & + \Big( {\cos(y) \over \sin(y)}  + K^2 { \cos(x^\prime) \over
 \sin(x^\prime) } \Big) \Big( { {\partial A} \over {\partial x} } ( y -
y^\prime )  + { {\partial B} \over {\partial x} } (x -x^\prime) \Big)
\\ \dot {x^\prime} &=& - \sin(x^\prime) \cos(y^\prime) \cos(z^\prime) -
K^2 \cos(x^\prime) \sin(y^\prime) \cos(z^\prime) - K^2 {\sin(x^\prime)
\over \sin(y^\prime)} \cos(z^\prime) (x - x^\prime) \\ \dot {y^\prime}
&=& - \cos(x^\prime) \sin(y^\prime) \cos(z^\prime) + K^2 \sin(x^\prime)
\cos(y^\prime) \cos(z^\prime) + K^2 {\sin(y^\prime) \over
\sin(x^\prime)} \cos(z^\prime) (y - y^\prime) \\ \dot {z^\prime} &=& 2
\cos(x^\prime) \cos(y^\prime) \sin(z^\prime) + \mu + \nu + { { \partial
\Ha 2 } \over { \partial y^\prime } }  K^2 {\rm cosec}^2(x^\prime)
 ( y - y^\prime) -  K^2 { { \partial \Ha 2 } \over { \partial x^\prime
} }{\rm cosec}^2(y^\prime) ( x - x^\prime ) \nonumber \\
 &  & + \alpha { {\partial A} \over {\partial y^\prime} } ( y -
 y^\prime ) + \alpha { {\partial B} \over {\partial y^\prime} } ( x -
x^\prime ) - \beta { {\partial A} \over {\partial x^\prime} } ( y -
y^\prime ) - \beta { {\partial B} \over {\partial y^\prime} } ( x -
x^\prime ) \end{eqnarray} where \begin{eqnarray} \mu &= K^2 {
\cos(y^\prime) \over \sin(y^\prime) }& \Big[ -\sin(x^\prime)
\cos(y^\prime) \sin(z^\prime) + \sin(x) \cos(y) \sin(z) \nonumber \\
  &                &\times { \sin(x) \over \sin(x^\prime) } \Big( { {
\cos(x^\prime) \sin(y^\prime) - K^2 \sin(x^\prime) \cos(y^\prime) }
\over { \cos(x) \sin(y^\prime) - K^2 \sin(x) \cos(y^\prime) } } \Big)
\Big] \nonumber  \\ \nu &= K^2 { \cos(x^\prime) \over \sin(x^\prime) }&
\Big[ \cos(x^\prime) \sin(y^\prime) \sin(z^\prime) - \cos(x) \sin(y)
\sin(z) \nonumber \\
  &                &\times { \sin(y) \over \sin(y^\prime) } \Big( { {
\sin(x^\prime) \cos(y^\prime) + K^2 \cos(x^\prime) \sin(y^\prime) }
\over { \cos(y) \sin(x^\prime) + K^2 \sin(y) \cos(x^\prime) } } \Big)
\Big] \nonumber \end{eqnarray} \begin{eqnarray} \alpha &=& - {
\cos(x^\prime) \over \sin(x^\prime) } +
 K^2 { \cos(y^\prime) \over \sin(y^\prime) } \nonumber \\ \beta &=&  {
\cos(y^\prime) \over \sin(y^\prime) } +
 K^2 { \cos(x^\prime) \over \sin(x^\prime) } \nonumber \end{eqnarray}

We now notice that on the subspace defined by equation
(\ref{sectplane}) $ \dot x = \dot {x^\prime}, \dot y = \dot
{y^\prime}, \dot z = \dot {z^\prime} $.  Thus the subspace is
invariant.

Moreover in this subspace the vector field is \begin{eqnarray} \dot x
&=& - \sin(x) \cos(y) \cos(z) - K^2 \cos(x) \sin(y) \cos(z) \nonumber
\\ \dot y &=& - \cos(x) \sin(y) \cos(z) + K^2 \cos(y) \sin(x) \cos(z)
\nonumber \\ \dot z &=& 2 \cos(x) \cos(y) \sin(z) \nonumber
\end{eqnarray} which are precisely the equations governing the
Chandrashekhar flow~\cite{HandK} where $(x,y,z)$ are coordinates of
the fluid particles in the Lagrangian description applicable to
Rayleigh-Benard convection with rotation and $K$ is constant
proportional to rotation of the fluid~\cite{HandK}.  
%
%
\subsubsection{ Arter flow }

Now we show that the Arter flow admits the description in our framework
as follows:

Consider Nambu manifold $(\R^6,\omegath)$. Let $\{ x, y, z, x^\prime,
y^\prime, z^\prime\}$ be the coordinates on $\R^6$ and $\omegath = dx
\wedge dy \wedge dz + dx^\prime \wedge dy^\prime \wedge dz^\prime$ in
this coordinate. Let $\Ho$ and $\Ht$ be the N ambu functions defined as
\begin{eqnarray} \Ho &=& \log \Big( { {\sin(x)} \over {\sin(y)}} \Big)
- \log \Big( { {\sin(x^\prime)} \over {\sin(y^\prime)}} \Big) - 2b
{{\cos(y) \cos(2z^\prime)} \over {\sin(x^\prime) \cos(z)}} (x -
x^\prime) + 2b {{\cos(x) \cos(2z^\prime)} \over {\sin(y^\prime) \cos(z
)}} (y - y^\prime)\\ \Ht &=& \sin(x) \sin(y) \sin(z) - \sin(x^\prime)
\sin(y^\prime) \sin(z^\prime) - (x - x^\prime)C + (y - y^\prime) D
\end{eqnarray} where \begin{eqnarray} C &=& { {\cos(x) \sin(y) \Big( (
\cos^2(x) + \cos^2(y) ) \cos(2z) \sin(z) - (\cos(2x) \cos(2y)) \sin(2z)
\cos(z) \Big) } \over { \cos(2z) ( \cos^2(x) - \cos^2(y))}} \nonumber
\\ D &=& { {\cos(y) \sin(x) \Big( ( \cos^2(x) + \cos^2(y) ) \cos(2z)
\sin(z) - (\cos(2x) \cos(2y)) \sin(2z) \cos(z) \Big) } \over { \cos(2z)
( \cos^2(x) - \cos^2(y))}} \nonumber \end{eqnarray} where $b$ is
constant. We have omitted the rather lengthy equations. As in the case
of Chandrashekhar flow, the subspace defined by \begin{equation} x =
x^\prime, y = y^\prime, z = z^\prime \label{sectplane1} \end{equation}
is an invariant subspace. The equations of motion for $(x,y,z)$ ( or
equivalently $(x^\prime,y^\prime,z^\prime)$) in this subspace are
\begin{eqnarray} \dot x &=& - \sin(x) \cos(y) \cos(z) + b \cos(2x)
\cos(2z) \nonumber \\ \dot y &=& - \cos(x) \sin(y) \cos(z) + b \cos(2y)
\cos(2z) \nonumber \\ \dot z &=& 2 \cos(x) \cos(y) \sin(z) - b
(\cos(2x) +\cos(2y))\sin(2z) \nonumber \end{eqnarray} which are
precisely the equations governing the Arter flow~\cite{HandK} where
$(x,y,z)$ are coordinates of the fluid particles in the Lagrangian
description applicable to Rayleigh-Benard convection with
rotation~\cite{HandK}.  
%
%
\subsection {Coupled rigid bodies }

We now consider the simplest case of a coupling between two symmetric
tops. The coupling introduced is proportional to the z component of
angular momenta of each rotor (Such an idealized situation corresponds
under certain assumptions to the case of two s ymmetric tops
interacting with each other through magnetic moment coupling). The
equations of motions for the angular momenta of the tops in their
respective body coordinates are

\begin{eqnarray} \dot{x_1} &=& {1 \over {I_{y_1} I_{z_1}}} [ y_1 z_1
(I_{z_1} - I_{y_1}) + I_{z_1} C_3 y_1 z_2 ] \nonumber \\ \dot{y_1} &=&
- {1 \over {I_{x_1} I_{z_1}}} [ x_1 z_1 (I_{z_1} - I_{x_1}) + I_{z_1}
C_3 x_1 z_2 ] \nonumber \\ \dot{z_1} &=& 0 \nonumber \\ \dot{x_2} &=&
{1 \over {I_{y_2} I_{z_2}}} [ y_2 z_2 (I_{z_2} - I_{y_2}) + I_{z_2} C_3
y_2 z_1 ] \nonumber \\ \dot{y_2} &=& - {1 \over {I_{x_2} I_{z_2}}} [
x_2 z_2 (I_{z_2} - I_{x_2}) + I_{z_2} C_3 x_2 z_1 ] \nonumber \\
\dot{z_2} &=& 0 \nonumber \end{eqnarray}

These equations have the generalized Nambu form in the sense of this
paper and can be obtained from the Nambu functions.

\begin{eqnarray} {\Ho } &=& {\div 1 2 } ( {x_1}^2 + {y_1}^2 + {z_1}^2 )
+ {\div 1 2 } ({x_2}^2 +{y_2}^2 + {z_2}^2 ) \nonumber \\ {\Ht } &=&
{\div 1 2 } {\Big(} { \div {{x_1}^2} {I_{x_1}} }
	   + { \div {{y_1}^2} {I_{y_1}} } + { \div {{z_1}^2} {I_{z_1}}
	   } {\Big)} + {\div 1 2 }{ \Big(} { \div {{x_2}^2} {I_{x_2}} }
	   + { \div {{y_2}^2} {I_{y_2}} } + { \div {{z_2}^2} {I_{z_2}}
	   }{ \Big) } + { C_3 {{z_1} {z_2}}  } \nonumber
\end{eqnarray}

It is obvious that the constant $C_3$ depends on the initial
orientation in lab frame. In the absence of coupling the tops obey
Euler equations individually. In above vector field the terms like
$I_{z_1} C_3 y_1 z_2$ can be considered as effective externa l torque
due to presence of another body. The important point to note is that
this torque just changes the precession frequency of the rigid body.
%
%
\section{ Conclusions }

We have developed a geometric framework for the formulation of
generalized Nambu systems. This formalism is more suitable from the
view point of dynamical systems. As demonstrated with the example of
Arter flow, a possibly non-integrable flow finds a desc ription in
terms of generalized Nambu flow. Specifically, the Chandrashekhar flow
and the Arter flow have been identified with the motion, which takes
place in an invariant three dimensional subspace of a six dimensional
Nambu system.

An interesting feature of the present formalism is the following.
Whereas a three bracket of functions gives rise to a non-associative
structure, a Nambu Poisson bracket of 2-forms gives rise to a Lie
algebra. It was found that formulations involving 2-fo rms provide a
natural approach to a Nambu system of order three. We feel that it is
worth investigating further the issues such as symmetries, reduction
and integrability for such systems. Nambu systems of higher order could
also be investigated. However,
 so far we have not carried this out, for the want of appropriate
 physical examples.
%
%
\vskip 0.125in
\centerline {\bf ACKNOWLEDGMENT}
\vskip 0.125in

We thank Dr. Hemant Bhate for the critical reading of the manuscript
and for extensive help in all aspects. We also thank Prof. K. B.
Marathe for his comments and for discussions.  We thank Ashutosh Sharma
for pointing out ref.~\cite{HandK}, Prof. N. Mukunda for useful
discussion and M. Roy for comments. One of the authors (SAP) is
grateful to CSIR (India) for financial assistance and the other (ADG)
is grateful to UGC (India) for
 financial assistance at the initial stages of work.
%
%
\vskip 0.125in
\centerline {\bf APPENDIX}
\vskip 0.125in

\begin{defn} (Block diagonal form) : Let $(\M, \omegath)$ be a Nambu
manifold. A 2-form $\a$ is called \underline{block diagonal form} if
for some $X \in \X(\M)$ \begin{eqnarray} i_X \omegath = \a \nonumber
\end{eqnarray} \end{defn}

\begin{defn} (Non diagonal form) : Let $(\M, \omegath)$ be a Nambu
manifold. A 2-form $\a$ is called \underline{non diagonal form} if for
each $z_1, z_2 \in \X(\M)$ such that $\a(z_1, z_2) \not= 0$
$\not\exists$ $X$ with the property $\omegath (X, z_1, z_ 2) = \a(z_1,
z_2)$.  \end{defn} {\bf Remarks:} \begin{enumerate} \item A form is not
non diagonal form does not imply that it is block diagonal.
\end{enumerate} \begin{prop} Let $(\M, \omegath)$ be a Nambu manifold,
let $\a$ be any 2-form then \begin{eqnarray} \a = \a^d + \a^\prime
\nonumber \end{eqnarray} where $\a^d$ is block diagonal form and
$\a^\prime$ is non diagonal form, and the decomposition is unique.
\end{prop} \pf Consider Darboux coordinates \begin{eqnarray} \a &=&
\sum_{i,j = 0}^{n-1} \sum_{l,m = 1}^3 \a_{3i+l\;\;3j+m} dx^{3i+l}
\wedge dx^{3j+m} \nonumber \\ &=& \sum_{i=0}^{n-1} \sum_{l,m=1}^3
\a_{3i+l\;\;3i+m} dx^{3i+l} \wedge dx^{3i+m} \nonumber \\
 &+& \sum_{i,j = 0, i \not= j}^{n-1} \sum_{l,m=1}^3 \a_{3i+l\;\;3j+m}
 dx^{3i+l} \wedge dx^{3j+m} \nonumber \end{eqnarray} It is easy to see
that the first term which we denote by $\a^d$ is block diagonal form
and the second term which we denote by $\a^\prime$ is non diagonal
form. Now we prove the uniqueness of the decomposition. Let $\a_1^d,
\a_1^\prime$ and $\a_2^d, \a_2^\prime$ are two distinct decompositions
of $\a$ such that $\a_1^d, \a_2^d$ are block diagonal forms and
$\a_1^\prime, \a_2^\prime$ are non diagonal forms. So we have $\a_1^d +
\a_1^\prime = \a_2^d + \a_2^\prime$ this implies that $\a_2\prime$ is
not non di agonal. hence $\a$ has unique decomposition.
\newline\rightline{$\Box$}

\begin{defn} We define a map \\ $\sharp : \Omega^0_2 (\M) \rightarrow
\X(\M) : \a \mapsto \a^\sharp$ such that \begin{eqnarray} i_{\a^\sharp}
\omegath = \a^d \nonumber \end{eqnarray} where $\a^d$ is block diagonal
part of $\a$ \end{defn} This map can be identified as the map
introduced in section~\ref{Nmanifold}. Since $\a^d$ is unique the
definition of the map is coordinate free.  
%
%

%
%
\end{document}